\def \SAIT #1 #2 {{\em Mem.\ Soc.\ Astron.\ It.\/} {\bf #1}, #2}
\def \MESS #1 #2 {{\em The Messenger\/} {\bf #1}, #2}
\def \ASTRNACH #1 #2 {{\em Astron. Nach.\/} {\bf #1}, #2}
\def \AAP #1 #2 {{\em Astron. Astrophys.\/} {\bf #1}, #2}
\def \AAL #1 #2 {{\em Astron. Astrophys. Lett.\/} {\bf #1}, L#2}
\def \AAR #1 #2 {{\em Astron. Astrophys. Rev.\/} {\bf #1}, #2}
\def \AAS #1 #2 {{\em Astron. Astrophys. Suppl. Ser.\/} {\bf #1}, #2}
\def \AJ #1 #2 {{\em Astron. J.\/} {\bf #1}, #2}
\def \ANNREV #1 #2 {{\em Ann. Rev. Astron. Astrophys.\/} {\bf #1}, #2}
\def \APJ #1 #2 {{\em Astrophys. J.\/} {\bf #1}, #2}
\def \APJL #1 #2 {{\em Astrophys. J. Lett.\/} {\bf #1}, L#2}
\def \APJS #1 #2 {{\em Astrophys. J. Suppl.\/} {\bf #1}, #2}
\def \APSS #1 #2 {{\em Astrophys. Space Sci.\/} {\bf #1}, #2}
\def \ASR #1 #2 {{\em Adv. Space Res.\/} {\bf #1}, #2}
\def \BAIC #1 #2 {{\em Bull. Astron. Inst. Czechosl.\/} {\bf #1}, #2}
\def \JSQRT #1 #2 {{\em J. Quant. Spectrosc. Radiat. Transfer\/} {\bf #1}, #2}
\def \MN #1 #2 {{\em Mon. Not. R. Astr. Soc.\/} {\bf #1}, #2}
\def \MEM #1 #2 {{\em Mem. R. Astr. Soc.\/} {\bf #1}, #2}
\def \PLR #1 #2 {{\em Phys. Lett. Rev.\/} {\bf #1}, #2}
\def \PASJ #1 #2 {{\em Publ. Astron. Soc. Japan\/} {\bf #1}, #2}
\def \PASP #1 #2 {{\em Publ. Astr. Soc. Pacific\/} {\bf #1}, #2}
\def \NAT #1 #2 {{\em Nature\/} {\bf #1}, #2}

\documentstyle{memsait}
\input epsf.sty
\begin{opening}
\title{INFRARED AND OPTICAL STUDY OF THE TYPE~Ia SN~1998bu IN M96 
\footnote{Talk given at workshop on {\it Future Directions of Supernova 
Research}, Assergi, Italy, 29 Sept. to 2 Oct. 1998.  To be published in 
the {\it Journal of the Italian Astronomical Society.}}}
\author{PETER MEIKLE$^1$, MIGUEL HERNANDEZ$^1$}
\institute{$^1$Blackett Laboratory, Imperial College, London, U.K.\\}
\date{} 
\end{opening}

\begin{document}

\oddpagefooter{}{}{} 
\evenpagefooter{}{}{} 
\ 
\bigskip

\begin{abstract}
The type~Ia SN~1998bu was discovered in the galaxy M96 on 9~May~1998.
It is one of the closest type~Ia events of modern times and good
temporal coverage has been achieved.  Moreover its distance is
well-determined.  We describe here some of the first-season
photometric and spectroscopic observations.  While SN~1998bu is
unusually highly reddened, we can still deduce that it was a typical
type~Ia event.  Preliminary analysis suggests that for an H$_0$ of
about 60~km~s$^{-1}$~Mpc$^{-1}$, its (de-reddened) peak luminosity,
light curve shape and spectroscopic behaviour confirms the
relationships found between these parameters in other classic type~Ia
supernovae.
\end{abstract}

\section{Introduction}
SN~1998bu in the Sab galaxy M96 (NGC~3368) was discovered by Mirko
Villi (1998) on May~9.9~UT at a magnitude of about 13, 10~days before
maximum blue light, t$_{Bmax}$=May~19.7$\pm$0.5~UT (see below).  The
supernova was identified by Ayani {\it et al.} (1998) and Meikle {\it
et al.}  (1998) as being of type~Ia.  A prediscovery observation on
May~3.14~UT was reported by Faranda \& Skiff (1998).  This was about
16.5~days before t$_{Bmax}$.  This makes it one of the earliest ever
observations of a type~Ia supernova (A. Riess, private communication).
Theoretical models indicate a rise time to t$_{Bmax}$ of 16--18 days
(H{\"o}flich {\it et al.} 1998).  Faranda's point constrains
this to 16.6--18 days for SN~1998bu.  We therefore estimate an
explosion date of May~2.0$\pm$1.0~UT.  The Faranda \& Skiff
measurement was made with an unfiltered CCD, and converts to a
V~magnitude of +16.70$\pm$0.10 (A. Riess, private communication). \\

An important aspect of the discovery of this supernova is that its
parent galaxy, M96, already had an HST-Cepheid-determined distance
of 11.6$\pm$0.8~Mpc (Tanvir {\it et al.} 1995).  More recently,
using PN distances Feldmeier {\it et al.} (1997) obtained an even closer
distance of 9.6$\pm$0.6~Mpc.  However, for ease of comparison with
other type~Ia supernovae with HST-Cepheid distances, we shall adopt
the value of Tanvir {\it et al.}.  SN~1998bu is one of the closest
type~Ias of modern times, as well as being one of the earliest ever 
observed. \\

\section{Light Curves}
Good coverage was achieved by observers at the IAC80 (Tenerife) and
the JKT (La Palma).  Some data were also provided by the WIYN at Kitt
Peak.  Relative magnitude light curves are shown in Figure~1.  They
show the supernova magnitude relative to a comparison star
(GSC~00849--00931) on the same CCD frame. \\

\begin{figure}
\vspace{12cm}  
\includegraphics{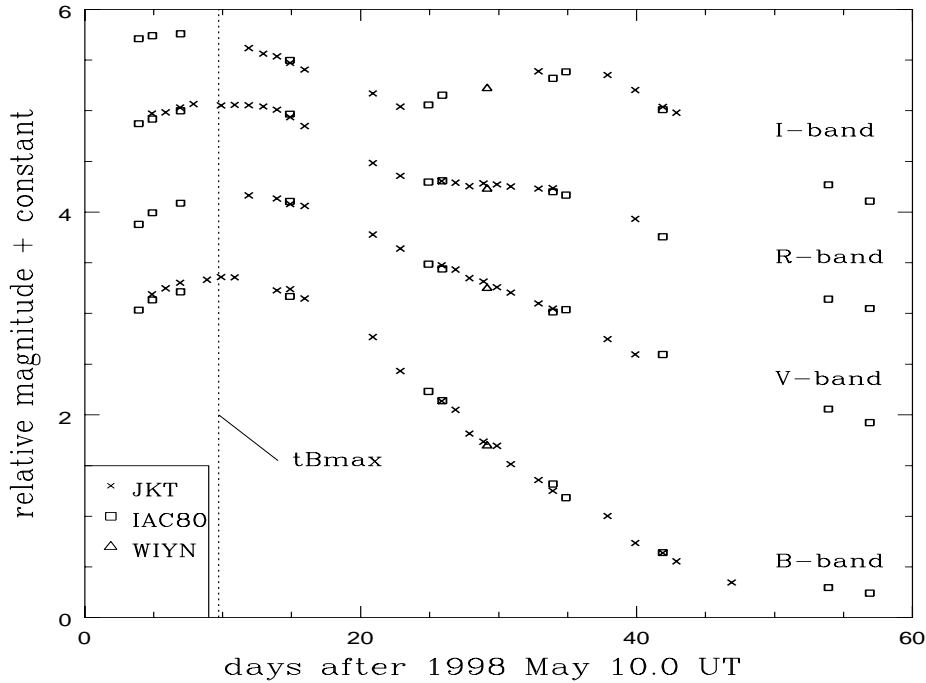}
\vspace{-5.5cm} 
\epsfysize=12cm 
\caption[h]{Relative magnitude optical light curves for SN~1998bu. 
For clarity they have been vertically displaced by arbitrary amounts.  
The epoch of maximum blue light, t$_{Bmax}$, corresponds to 1998~May~19.7~UT.}
\end{figure}

The epochs of maximum light were estimated by comparison with the
templates of Leibundgut (1988) and Schlegel (1995) (Figure~2).
Maximum light in the B-band was on May~19.7$\pm$0.5 days.  Henceforth,
this epoch will be adopted as the fiducial t$_{Bmax}$=0~days. The
V-band fluxes peaked at $\sim$+1~d, while the R and I-band fluxes
peaked at --3 to --4~d i.e. {\it before} t$_{Bmax}$.  This behaviour
has been noted in other type~Ias such as SN~1990N and SN1992A
(Suntzeff 1993, Leibundgut 1998, Lira {\it et al.} 1998).  It follows
from the fact that, even at this relatively early phase, the type~Ia
photosphere cannot be regarded as a simple black body.  The extent to
which the photosphere is not a black body becomes clear when we
inspect contemporary spectra (see below).  We also see a pronounced
secondary maximum in I at about +25~d together with an inflection in
R.  Again, this behaviour has been seen in other type~Ias (Schlegel
1995).  Using Landolt standards (Landolt 1992), we deduce a
preliminary value for $m_{Vmax}$ of +11.89$\pm$0.04. \\

SN~1998bu has yielded one of the earliest sets of {\it near-IR}
photometry ever obtained.  Indeed, this is the first time that IR
photometry for a normal type~Ia event has been acquired {\it before}
t$_{Bmax}$.  IR photometry obtained at the OAN (Mayya {\it et al.}
1998), TCS, IRTF, UKIRT and WHT telescopes are shown in Figure~2.
Interestingly, the pre-t$_{Bmax}$ IR light-curve data peak at about
--5~d and thus continue the trend of pre-t$_{Bmax}$ maxima seen in the
R and I-bands.  Also shown in Figure~2 are the JHK-band template light
curves of Elias {\it et al.} (1985). (We have slightly truncated Elias
{\it et al.}'s original templates so that the earliest epoch of the
template corresponds to Elias {\it et al.}'s earliest observation.)
The position of these templates were fixed in the time axis by the
epoch of t$_{Bmax}$ and have only been shifted vertically to provide
the best match to the data.  While the IR observations after
$\sim$+10days are rather sparse, we can see that the data are
consistent with the Elias {\it et al.} templates. \\

The absolute peak magnitudes and reddening for SN~1998bu are discussed 
below. \\

\begin{figure}
\vspace{12cm}  
\includegraphics{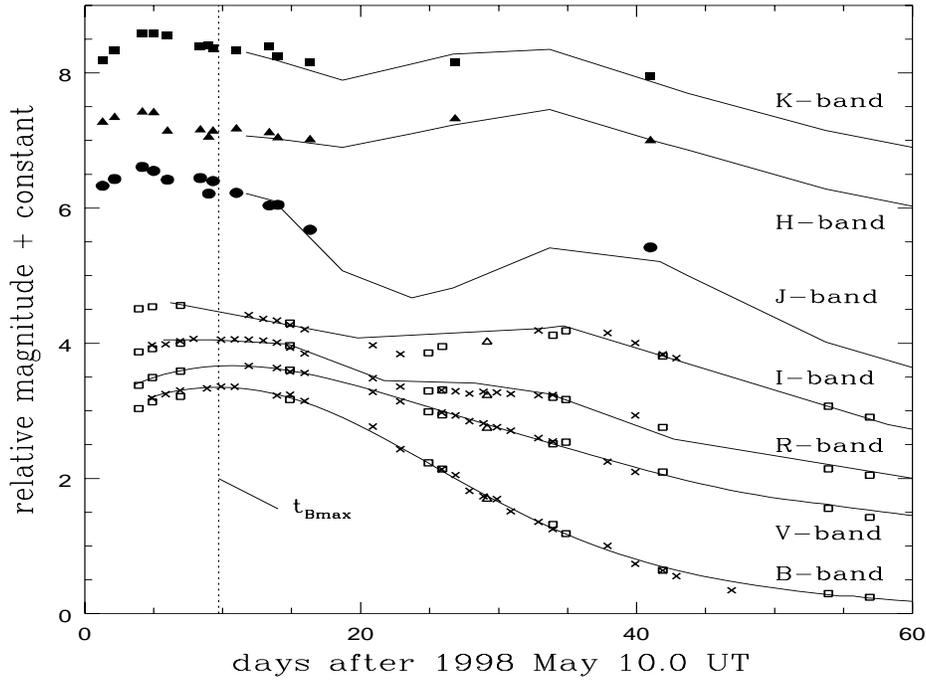}
\vspace{-5.5cm} 
\caption[h]{Relative magnitude infrared and optical light curves for
SN~1998bu.  For clarity they have been displaced vertically by
arbitrary amounts.  The IR photometry was obtained at the OAN 
(Mayya {\it et al.} 1998), TCS, IRTF, UKIRT and WHT telescopes.  Also
shown are template light curves in B \& V (Leibundgut 1988), R \&
I (Schlegel 1995) and J, H \& K (Elias {\it et al.}  1985). The BVR \&
I templates were shifted in both axes to give the best match to the
data.  The JHK templates were shifted only vertically.  Their
horizontal position was fixed by the epoch of t$_{Bmax}$ as indicated
in Elias {\it et al.}.}
\end{figure}

\begin{figure}
\vspace{11cm}  
\includegraphics{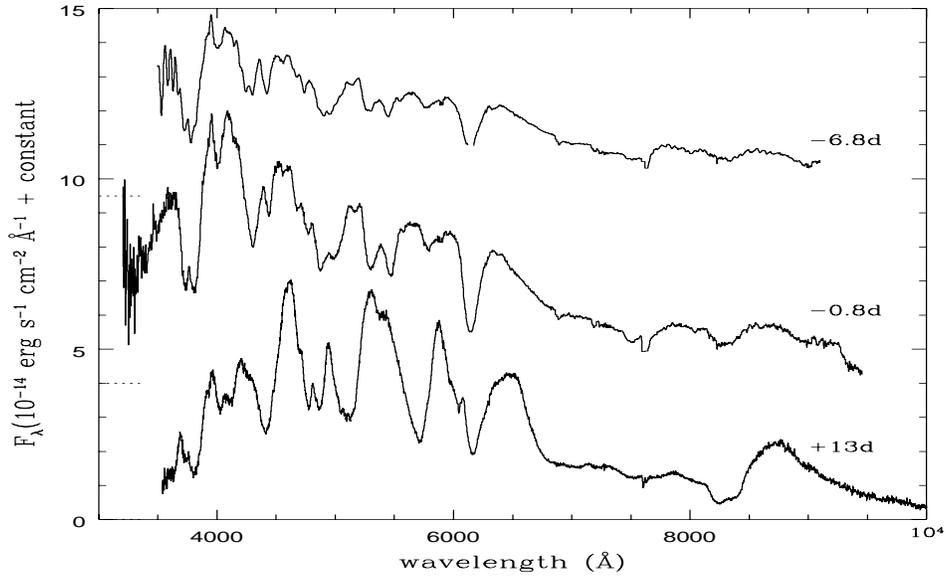}
\vspace{-5.5cm} 
\caption[h]{Optical spectra of SN~1998bu.  The epochs shown are with
respect to t$_{Bmax}$ = 1998~May~19.7.  The --6.8~d and --0.8~d
spectra have been shifted vertically for clarity.  Their zero flux
axes are indicated by the dotted lines. Fluxing is approximate only.
The --6.8~d spectrum was taken at the WHT, the --0.8~d spectrum at the
INT and the +13~d spectrum at the WIYN telescope.  The WIYN spectrum
is composed of a blue spectrum taken on May~31.2~UT and a red spectrum
taken on June~4.2~UT, scaled to match in their overlap region.  }
\end{figure}

\begin{figure}
\vspace{12cm}  
\includegraphics{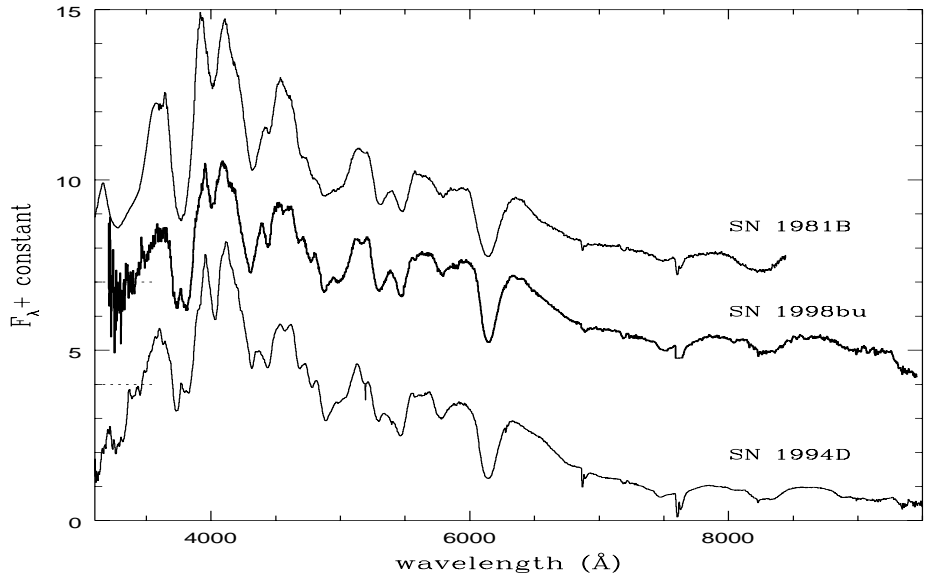}
\vspace{-5.5cm} 
\caption[h]{Spectra of three type~Ia supernovae close to t$_{Bmax}$.
The SN~1981B and SN~1998bu spectra have been shifted vertically and
scaled for clarity.  Their zero flux axes are indicated by the dotted
lines on the left hand axis.  The SN~1981B spectrum is courtesy of
B.~Leibundgut and P.~Nugent.  The SN~1994D spectrum is from Meikle
{\it et al.} (1996).}
\end{figure}

\section{Spectra}
Examples of the optical spectra are shown in Figure~3.  The --6.8~d
spectrum, taken with the ISIS spectrograph at the WHT (Meikle {\it et
al.} 1998) is the earliest reported.  The --0.8~d spectrum is from the
IDS at the INT.  The +13~d spectrum was obtained at the WIYN
telescope, Kitt Peak.  It is composed of a blue spectrum taken on 
May~31.2~UT and a red spectrum taken on June~4.2~UT.  The spectra
were scaled to match in their overlap region.  Inspection of Figure~3
immediately reveals the characteristic deep, broad Si~II absorption at
6120~\AA, as well as other absorption lines (at 5750, 5440, 5270, 4860, and
4270~\AA.) typical of a normal type~Ia supernova at early times.  The
expansion velocity derived from the shift of the Si~II absorption
minimun is about 11,300~km/s. \\

In Figure~4 we compare SN 1998bu at maximum light with the type~Ia
SNe~1981B and 1994D.  Apart from the greater degree of reddening
in the SN~1998bu spectrum, the three spectra are quite similar.
The spectra of SNe~1998bu and 1981B are particularly alike. \\

We were unable to obtain any early-time IR spectra as the IR
spectrograph at UKIRT (CGS4) was undergoing upgrading during May and
early June.  An IR spectrum was obtained on June~24.2~UT (t=+35.5~d),
spanning 8,300--25,000~\AA\ (Figure~5). As usual, we see the dramatic
drop in the 10,000 to 12,000~\AA\ region responsible for the red J--H colour
at this time.  The strong feature at about 8,700~\AA\ is due to the
calcium triplet.  There is a particularly prominent feature at
10,000~\AA.  Comparison with the IR spectrum of the type~Ia SN~1995D
(Bowers {\it et al.} 1997) at 92~d (Figure~6) suggests that it fades
quite rapidly.  P. H{\"o}flich (private communication) has recently
identified it as a very strong Fe~II line.  Many of the other features
in the IR spectrum are probably to doubly ionised cobalt and iron
(Bowers {\it et al.} 1997). \\

\begin{figure}
\vspace{12cm}  
\includegraphics{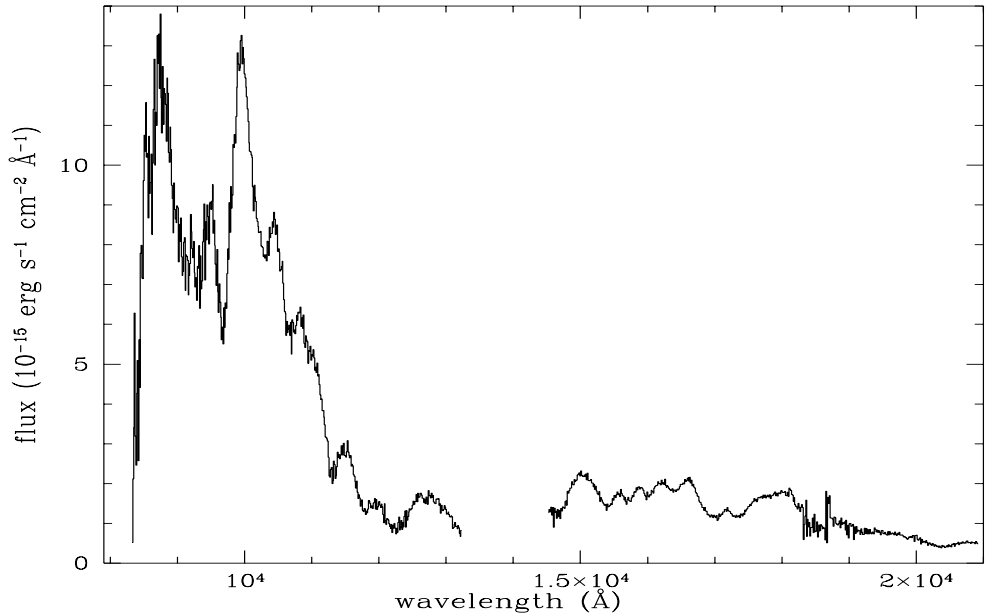}
\vspace{-5.5cm} 
\caption[h]{Infrared spectrum of SN~1998bu at +35.5~d, taken with the
CGS4 spectrograph at UKIRT.  Note the strong, relatively isolated
feature at 10,000~\AA, probably due to Fe~II.}
\end{figure}

\begin{figure}
\vspace{12cm}  
\includegraphics{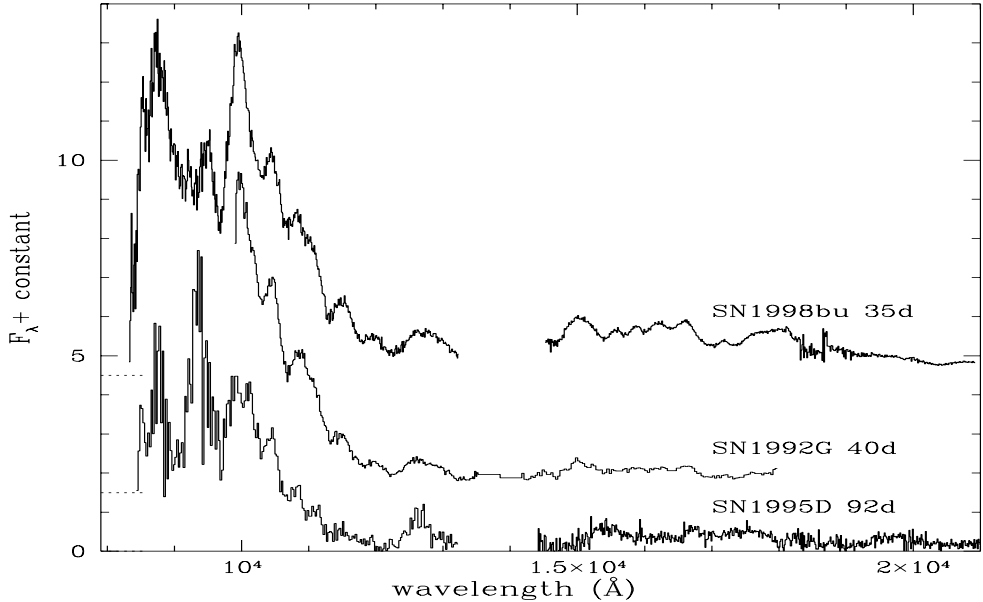}
\vspace{-6.3cm} 
\caption[h]
{Infrared spectra of three type~Ia supernovae illustrating the
spectral evolution.  The SN~1998bu and SN~1992G spectra have been
shifted vertically and scaled for clarity.  Their zero flux axes are
indicated by the dotted lines on the left hand axis.  The SNe~1992G \&
1995D spectra are from Bowers {\it et al.} (1997).  Note the weakening
of the 10,000~\AA\ feature by +92~d.}
\end{figure}

The lack of any early-time IR spectra for SN~1998bu meant that we do
not know if it exhibited the 10,500~\AA\ absorption feature seen in
SNe~1991T and 1994D (Meikle {\it et al.} 1996).  However, we did
obtain IR spectra for the type~Ia SN~1998aq at --8.5~d and --6.5~d.
In these the 10,500~\AA\ feature was clearly present but was of
somewhat smaller depth than in SN~1994D. \\

\section{Reddening and Absolute Peak Magnitude} 
As already mentioned, SN~1998bu exhibited an unusually high degree of
reddening. We consider two methods for estimating the extinction {\it
viz.} interstellar absorption and supernova colours.  Measurements of
interstellar NaI D absorption by Munari {\it et al.}  (1998) indicate
an E(B-V)=0.15 (EW=0.35~\AA) in the parent galaxy, with a further 0.06
(EW=0.19~\AA) arising in our own Galaxy.  Thus, we have A$_V$=0.65,
which is unusually high for a type~Ia supernova. (The median A$_V$ in
Mazzali {\it et al.} (1998) for 14 well-observed type~Ias is 0.13).
However, Suntzeff {\it et al.} (1999) argue that the Munari {\it et
al.} estimate is actually an {\it upper} limit and so is inconsistent
with extinction derived using supernova colours. \\

The other approach is to use the supernova colours themselves and
compare them with those expected for an unreddened type~Ia supernova
of the same sub-type.  Given the high degree of similarity in the
spectral features of SNe~1998bu and 1981B, we assumed that the two
supernovae are intrinsically identical.  We then used the empirical
reddening law of Cardelli {\it et al.} (1989) and gradually reddened
the SN~1998bu optical spectrum to provide the best match to the
SN~19981B spectrum.  From this we derive a relative reddening of
E(B-V)=0.26$\pm$0.02 for SN~1998bu.  For the total reddening to
SN~1981B we adopt E(B-V)=0.138$\pm$0.020 (Suntzeff {\it et al.}
1999). \\

After de-reddening, we compared our SN~1998bu observations with the
light curve studies of Phillips (1993) and Riess {\it et al.} (1995)
and with the spectroscopic sequence study of Nugent {\it et al.}
(1996).  We find that for SN~1998bu to match the trends indicated by
these studies it is necessary to adopt an H$_0$ of
55--60~km~s$^{-1}$~Mpc$^{-1}$.  Using the Phillips {\it et al.} (1999)
relation relating M$_{Vmax}$ with the decline rate parameter
$\Delta$m$_{15}$(B), we obtain an H$_0$ of
61$\pm$6~km~s$^{-1}$~Mpc$^{-1}$, which is consistent with Suntzeff
{\it et al.}'s (1999) V-band-derived value of
64.7$\pm$7.6~km~s$^{-1}$~Mpc$^{-1}$.  However, we emphasize that our
result is preliminary.  A more comprehensive description and analysis
of this work is in preparation.

\acknowledgements
We are grateful to the many observers who contributed some of their
scheduled telescope time to observing SN~1998bu.  In particular, we
thank Chris Benn, Tom Geballe, Di Harmer, Pete Hammersley, Simon Kemp,
Don Pollacco, Nic Walton and Bill Vacca for their assistance in
acquiring these observations.  We also thank David Branch, Peter
H{\"o}flich, Paolo Mazzali, Phil Pinto, Adam Riess, Nick Suntzeff and
Craig Wheeler for helpful discussions.


\end{document}